%% file: main.tex
\documentclass{article}
\usepackage{spconf,amsmath,graphicx}

\usepackage{comment}
\usepackage{amssymb}
\usepackage{xcolor}
\usepackage{hyperref}
\usepackage{multirow}
\usepackage{makecell}
\usepackage{bm}
\usepackage{amsmath, amsfonts}
\usepackage{subcaption,booktabs}
\usepackage{makecell}
\usepackage{pifont}
\newcommand{\xmark}{\ding{54}}
\newcommand\Tstrut{\rule{0pt}{2.6ex}}
\usepackage{hhline}

\pdfoutput=1
\usepackage{colortbl}
\usepackage{subcaption}

\newcolumntype{?}{!{\vrule width 1.3pt}}

\title{SpeechPrompt v2: Prompt Tuning for Speech Classification Tasks}
\name{Author(s) Name(s)\thanks{Thanks to XYZ agency for funding.}}
\address{Author Affiliation(s)}
\name{%

\vspace{-0.5em}
\begin{tabular}{@{}c@{}}
Kai-Wei Chang$^{1\star}$ \qquad 
Yu-Kai Wang$^{2\star}$ \qquad 
Hua Shen$^{3}$ \qquad 
Iu-thing Kang$^{4}$ \qquad \\
Wei-Cheng Tseng$^{1}$ \qquad 
Shang-Wen Li$^{5}$ \qquad
Hung-yi Lee$^{1}$
\thanks{$^*$ The first two authors contribute equally.}
\end{tabular}
}

\address{$^1$Graduate Institute of Communication Engineering, National Taiwan University, Taiwan\\$^{1,2,4}$National Taiwan University, Taiwan, $^3$Pennsylvania State University, $^5$Meta AI}
     

\begin{document}
%
\maketitle

\input{0_abstract.tex}
\section{Introduction}
\label{sec:intro}
\input{1_introduction.tex}

\section{Related Works}
\label{sec:related}
\input{2_related.tex}

\section{Method}
\label{sec:method}
\input{3_method.tex}

\input{assets/main_result_table}

\section{Experimental Setup}
\label{sec:experiment}
\input{4_experiment.tex}

\section{Results}
\label{sec:majhead}
\input{5_result.tex}

\section{Conclusions}
\label{sec:print}
\input{6_conclusion.tex}

\section{Acknowledgement}
\input{7_acknowledgement.tex}


\bibliographystyle{IEEEbib-abbrev}
\bibliography{refs}

\end{document}

%% file: 0_abstract.tex
\begin{abstract}
Prompt tuning is a technology that tunes a small set of parameters to steer a pre-trained language model (LM) to directly generate the output for downstream tasks. Recently,  prompt tuning has demonstrated its storage and computation efficiency in both natural language processing (NLP) and speech processing fields. These advantages have also revealed prompt tuning as a candidate approach to serving pre-trained LM for multiple tasks in a unified manner. For speech processing, SpeechPrompt shows its high parameter efficiency and competitive performance on a few speech classification tasks. However, whether SpeechPrompt is capable of serving a large number of tasks is unanswered. In this work, we propose SpeechPrompt v2, a prompt tuning framework capable of performing a wide variety of speech classification tasks, covering multiple languages and prosody-related tasks. The experiment result shows that SpeechPrompt v2 achieves performance on par with prior works with less than 0.15M trainable parameters in a unified framework. The SpeechPrompt project website is at \url{https://ga642381.github.io/SpeechPrompt}.
\end{abstract}
\begin{keywords}
prompt tuning, self-supervised learning, spoken language model, speech classification
\end{keywords}

%% file: 1_introduction.tex
Recently, self-supervised pre-trained models have become an important component of speech processing. By utilizing structural knowledge learned from large-scale unlabeled corpus, these models can generate general-purpose representations that benefit a variety of speech processing tasks~\cite{mohamed2022self}. When leveraging a self-supervised speech model for a downstream task of interest, a typical way is to follow the \textbf{``pre-train, fine-tune'' paradigm}. Under this paradigm, a task-specific downstream model is built on top of the pre-trained self-supervised speech model. Then both models will be jointly fine-tuned with the objective function of the downstream task. Several works have demonstrated that this paradigm can achieve state-of-the-art performance on various speech processing tasks. However, this paradigm actually requires a delicately-designed downstream model and loss function for each task, causing an increasing burden of human labor. Also, it requires heavy computation and storage resources when the number of downstream tasks scales up.

\begin{figure}[t]
    \centering
    \includegraphics[width=0.90\columnwidth]{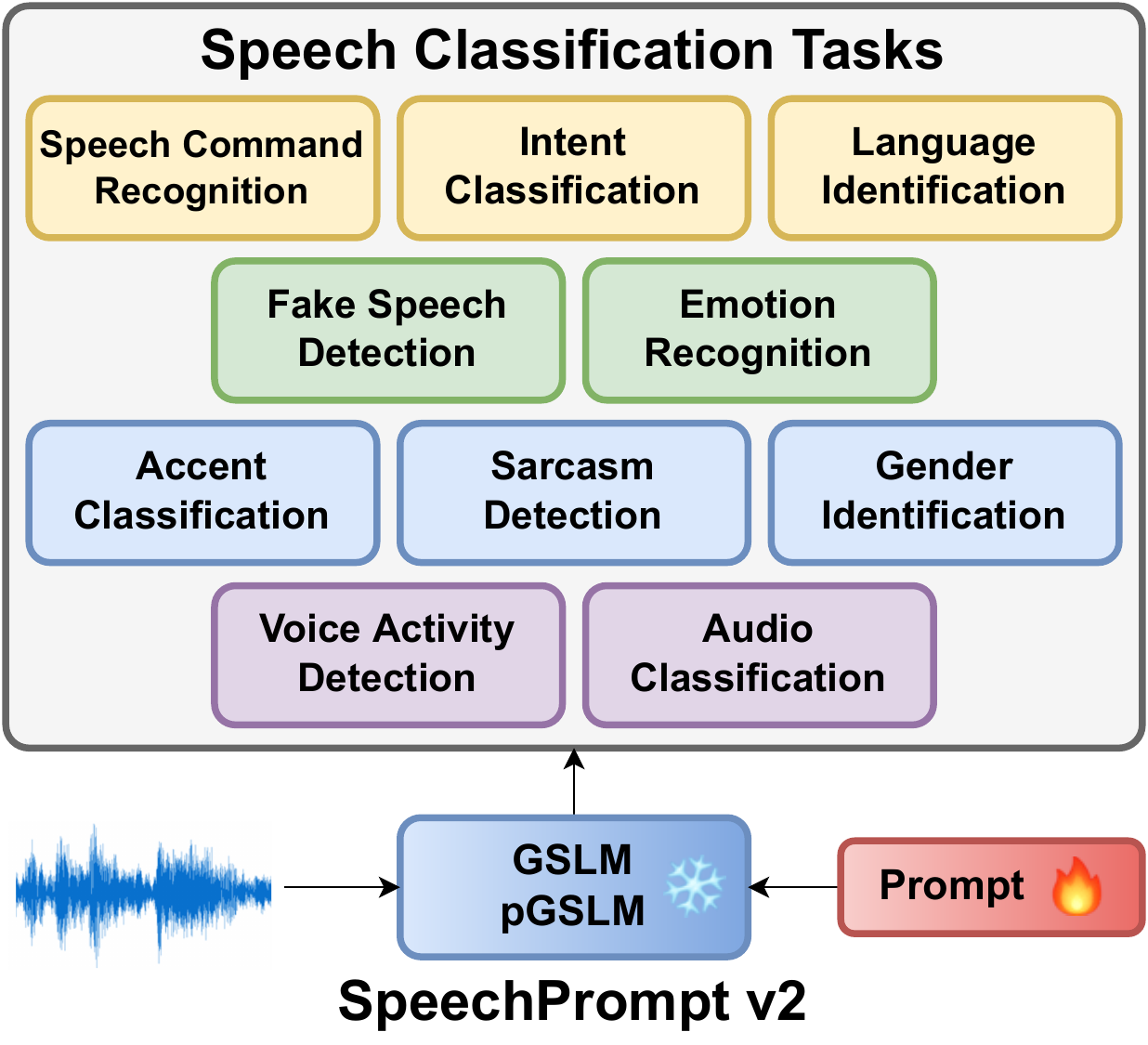}%
    
    \caption{The proposed SpeechPrompt v2. It demonstrates versatility and performs competitively in various speech classification tasks with minimal parameters updated. The pre-trained spoken LMs are frozen, while only the small set of prompt vectors are trainable.}
    \label{fig:framework}
\end{figure}

On the other hand, researchers explore the \textbf{prompting paradigm} \cite{liu2021pre} as an alternative method to utilize pre-trained models.
Prompting refers to the technique that finds a task-specific template or instruction to steer a pre-trained LM without modifying its architecture and parameters.
The prompting method can serve downstream tasks in a masked prediction or generative manner. For instance, in sentiment classification, an input sentence $\langle S \rangle$ can be fit into a template: ``$\langle S \rangle$. It was \_\_.'' and then fed into a pre-trained LM. The LM's output (e.g., great, terrible) is then transformed into sentiment classes (positive, negative) by a verbalizer, which is typically a hand-crafted mapping function. This approach enables us to determine the sentiment of $\langle S \rangle$.

Prompting offers several benefits over the typical ``pre-train, fine-tune'' paradigm. For instance, we no longer need to design a downstream model for each task. All tasks can be served in a unified framework, making it computationally and storage efficient. 
Under the premise of fixing the pre-trained LM, \textbf{prompt tuning} that updates a small number of continuous prompt vectors has been proposed as an efficient prompting method. For example, prefix-tuning~\cite{DBLP:conf/acl/LiL20} performs \emph{deep prompt tuning}, where trainable prompts are concatenated at the input of the LM's Transformer layers. 

SpeechPrompt~\cite{https://doi.org/10.48550/arxiv.2203.16773} applies prompt tuning on Generative Spoken Language Model (GSLM)~\cite{lakhotia2021generative} to perform speech classification and sequence generation tasks. However, it only shows the feasibility of prompting in limited speech processing tasks and datasets.
To this end, we would like to further examine the generalizability of the prompting paradigm on a wide range of speech classification tasks. We propose \textbf{SpeechPrompt v2}, a prompt tuning framework capable of performing various speech classification tasks, including content and prosody-related tasks, covering various languages. Our contributions are threefold:

\begin{itemize}
    \item We propose \emph{SpeechPrompt v2}, a prompt tuning framework capable of performing a wide range of speech classification tasks with performance comparable to prior works using uniformed model architectures and limited trainable parameters. 
    \item We introduce a collection of speech classification tasks to evaluate self-supervised spoken generative models. The tasks include content-related and prosody-related tasks in  multiple languages.
    \item We introduce a learnable verbalizer that consistently improves the performance for prompting spoken language models. Also, an analysis is conducted to verify its necessity.
\end{itemize}


%% file: 2_related.tex
\subsection{Generative Spoken Language Models}
Recently, generative \textbf{spoken LMs} have been proposed to model various speech characteristics, showing competitive performance in generating novel speech. \textbf{Generative Spoken Language Model (GSLM)}~\cite{lakhotia2021generative} performs generative language modeling on discrete units which serve as pseudo text carrying phonetic information encoded by SSL speech models. Moreover, \textbf{prosody-aware Generative Spoken Language Model (pGSLM)}~\cite{DBLP:conf/acl/KharitonovLPACL22} further extends the capability of GSLM by modeling the prosody in speech. A multi-stream Transformer was proposed to simultaneously model the discrete units, duration, and pitch information of the speech signal. These spoken LMs perform in a generative manner and can be regarded as speech versions of GPTs~\cite{DBLP:conf/acl/KharitonovLPACL22}, therefore offering an opportunity to perform prompting in the speech processing field.

\subsection{Prompting and Reprogramming in Speech Processing}
Originating from the NLP field, prompting has also been explored in the speech processing field. WAVPROMPT \cite{DBLP:conf/interspeech/GaoNQZCH22} adopts GPT-2 as the backbone LM and encodes audio into the prompt to perform spoken language understanding (SLU) tasks. SpeechPrompt \cite{https://doi.org/10.48550/arxiv.2203.16773} performs prompt tuning on GSLM for speech classification and sequence generation tasks. A heuristic frequency-based verbalizer is introduced to map the GSLM's vocabulary units to the classes of the downstream task. However, the frequency-based verbalizer does not significantly outperform the random-mapping method. To this end, we propose a learnable verbalizer in SpeechPrompt v2 to improve performance.

Another branch of utilizing a pre-trained model's capability for different tasks is \emph{model reprogramming}. In \cite{DBLP:conf/icml/YangTC21, DBLP:journals/corr/abs-2110-03894}, the input data (target domain) is first transformed with a task-specific function. The pre-trained acoustic model is then capable of generating labels for the reprogrammed data. These labels (source domain) are then mapped to the classes of downstream tasks (target domain) by a mapping function. This function serves the same role as the verbalizer in the prompting method and is usually a random mapping in the reprogramming literature.

%% file: 3_method.tex
We propose SpeechPrompt v2, a prompting framework for spoken LMs serving diverse downstream speech classification tasks. In general prompt tuning methods and SpeechPrompt v2, the pre-trained LMs are always fixed. Given a downstream task and the pre-trained spoken LM $\mathcal{M}$, we learn task-specific prompt vectors $\mathcal{P}$. By conditioning the LM $\mathcal{M}$ on the input $\bm{x}$ and the prompt vectors $\mathcal{P}$, we can generate the LM's output, which is then mapped to task labels using a verbalizer $v$ to make a prediction:
\begin{align}
    \bm{\hat{y}} = v(\mathcal{M}(\mathcal{P}; \bm{x})).
\end{align}
In SpeechPrompt v2, we adopted GSLM and pGSLM as our backbone LMs, and in the following, we explain the implementation details of SpeechPrompt v2 for bringing general prompting techniques to speech domains.

\subsection{Prompt Tuning}
In GSLM and pGSLM, an utterance is first encoded into discrete units $\bm{u}$ by an SSL model and a K-means quantizer. The difference between GSLM and pGSLM is that pGSLM also extracts duration $\bm{d}$ and pitch information $\bm{f}$ (log fundamental frequency) from the given utterance. 
Prompt vectors $\bm{p}^{I}$ are then concatenated at the input
\begin{align}
    \bm{x} = \left\{
    \begin{aligned}
    &[\bm{p}^I, \bm{e}(\bm{u})] &\text{ for GSLM} \\
    &[\bm{p}^I, \bm{e}(\bm{u}) + \bm{e}(\bm{d}) + \bm{e}(\bm{f})] &\text{ for pGSLM}        
    \end{aligned}
    \right.,
\end{align}
where $\bm{p}^{I}$ is trainable vectors in $\mathcal{P}$, and $\bm{e(\cdot)}$ denotes the input embedding of the spoken LMs.

In addition to concatenating the prompts at the input, we also adopt \emph{deep prompt tuning} \cite{DBLP:conf/acl/LiL20, https://doi.org/10.48550/arxiv.2203.16773} to enhance the prompt's instruction ability.
Given the $j^{th}$ Transformer layer that takes the embedding $\bm{x}^{j}=[\bm{x}^{j}_1,\bm{x}^{j}_2,\cdots,\bm{x}^{j}_T]$ as input, we concatenate the prompt vectors $\bm{p}^K$ and $\bm{p}^V$ to manipulate the key and the value in the attention function $Attn(Q, K, V)$ :
\begin{align}
K &= Concat(\bm{p}^K,\bm{x}^j_{l+1:T})W^K, \\
V &= Concat(\bm{p}^V,\bm{x}^j_{l+1:T})W^V,
\end{align}
where $\bm{p}^K$, $\bm{p}^V$ are trainable prompts in $\mathcal{P}$ and $l$ is the prompt length.

\subsection{Learnable Verbalizer}
\label{sec:learnable_verbalizer}
\begin{figure}[t]
    \centering
    \includegraphics[width=\columnwidth]{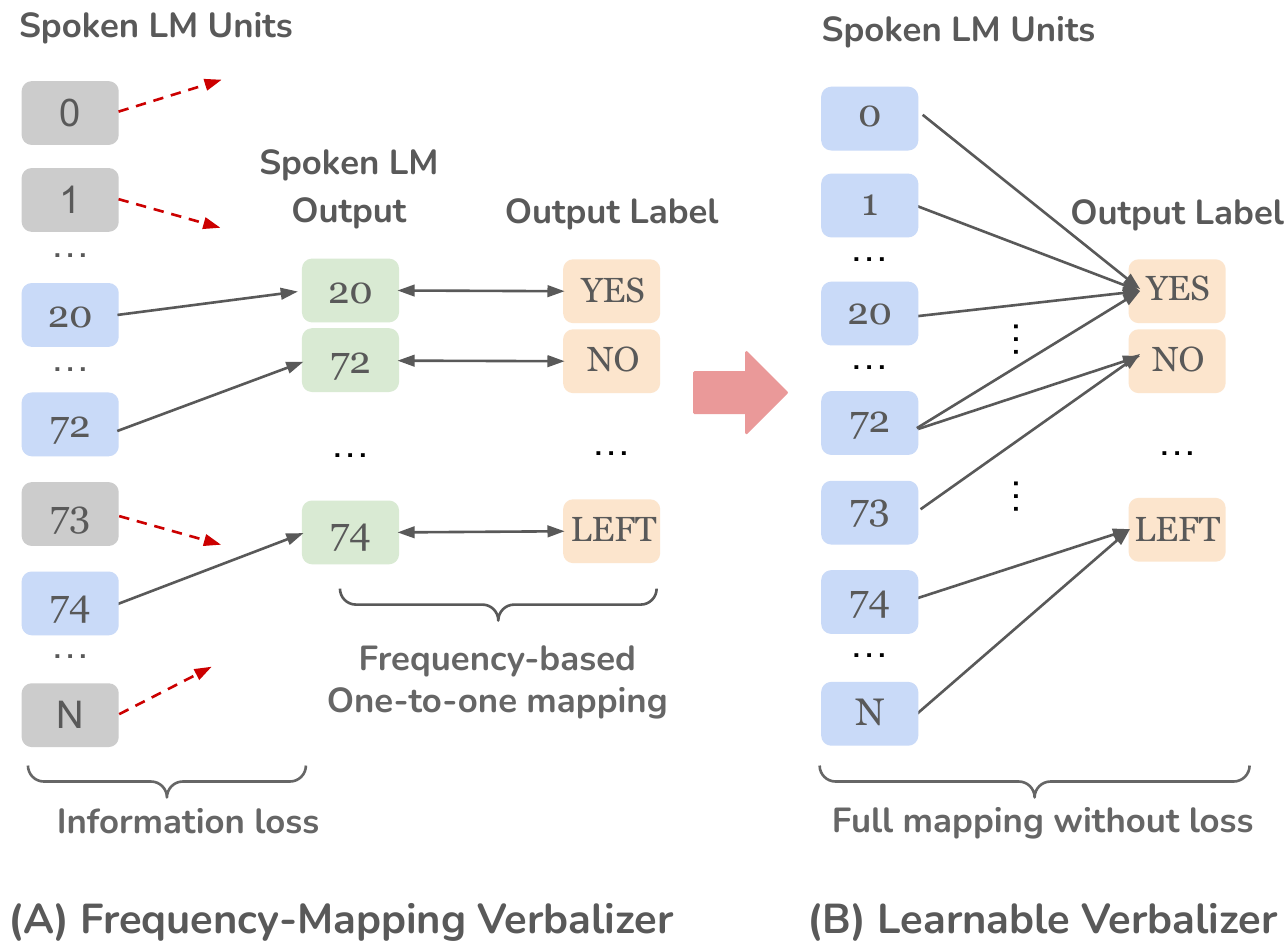}%
    \caption{The comparison between Frequency-Mapping Verbalizer (A) and the proposed learnable verbalizer in SpeechPrompt v2 (B).}
    \label{fig:verbalizer}
\end{figure}

To connect the LM's output to classes of downstream tasks, a verbalizer $v$ is adopted in the general prompting paradigm, and $v$ maps the LM's vocabulary to classes of tasks.
For example, in sentiment classification, the word ``great'' is mapped to the positive sentiment. 
SpeechPrompt~\cite{https://doi.org/10.48550/arxiv.2203.16773} leverages a statistical frequency-based mapping as the verbalizer without model estimation. However, this algorithm suffers from the information loss problem (Fig. \ref{fig:verbalizer}) and does not consistently outperform random mapping method.
In SpeechPrompt v2, we propose using a learnable linear model to implement the verbalizer $v$. $v$ takes as input the output distribution over the vocabulary of spoken LMs and maps the distribution to the classes of downstream tasks. Therefore, the trainable prompt vectors $\mathcal{P}$ and the verbalizer $v$ are optimized jointly with a loss function $\mathcal{L}$:
\begin{align}
    \mathcal{P}, v = \mathop{\arg\min}_{\mathcal{P}, v} \mathcal{L}(v(\mathcal{M}(\mathcal{P}; \bm{x})), \bm{y}),
\end{align}
where $\bm{y}$ denotes the ground truth of labels for downstream tasks. In SpeechPrompt v2, we use cross-entropy for every task.

%% file: assets/main_result_table.tex
\definecolor{mygreen}{HTML}{009B55}
\definecolor{myred}{HTML}{ED1B23}
\begin{table*}[!ht]
    \centering
    \caption{SpeechPrompt v2 on a wide variety of speech classification tasks. ``+'' denotes adopting a learnable verbalizer.
    For each task, we also list the performance of previous SOTA that are in "Fully supervised learning" or "Pre-train, Fine-tune" paradigms. These works serve as a performance topline that adopts the same dataset. The number in the parentheses shows the performance differences between SpeechPrompt and SOTA.
    To compare with the previous works, results are reported in terms of equal error rate (EER) for FSD, F1-score (F1) for SD and GID, and accuracy (ACC) for the remaining tasks.}
    \resizebox{\textwidth}{!}{\begin{tabular}{c c c c c c c c c c c }
        \toprule
        \cellcolor[HTML]{DAE8FC}{\textbf{Task}}  & \cellcolor[HTML]{DAE8FC}{\textbf{Metric}}                 & \cellcolor[HTML]{DAE8FC}{\textbf{Dataset}}       &    \cellcolor[HTML]{DAE8FC}{\textbf{Language}}   & \cellcolor[HTML]{DAE8FC}{\textbf{\#Class}}   & \cellcolor[HTML]{DAE8FC}{\textbf{SOTA (Topline)}}    & \cellcolor[HTML]{F5DED0}{\textbf{GSLM}}    & \cellcolor[HTML]{F5DED0}{\textbf{GSLM+}}   & \cellcolor[HTML]{F5DED0}{\textbf{pGSLM}}   & \cellcolor[HTML]{F5DED0}{\textbf{pGSLM+}}  \\ 
        \midrule
        \midrule
        \multirow{4}{*}{~\quad SCR \quad } & \multirow{4}{*}{ACC ($\uparrow$)}       & Google SC v1            &  En     & 12                     & 98.6 \cite{vygon2021learning} & 94.5 & 
                                               94.6 & 94.3 & \cellcolor[HTML]{EFEFEF}{\textbf{94.7 } (\textcolor{mygreen}{-3.9})} \\ \hhline{~~*{8}{-}} \Tstrut & \Tstrut 
                                               & Grabo SC              &  Du      & 36                     & 98.9 \cite{DBLP:conf/interspeech/TianG20} & 92.4 & \cellcolor[HTML]{EFEFEF}{\textbf{92.7 }(\textcolor{mygreen}{-6.2})} & 17.5 & 19.6 \\ \hhline{~~*{8}{-}} \Tstrut & \Tstrut 
                                               & Lithuanian SC         &  Lt    & 15                     & 91.8 \cite{DBLP:journals/corr/abs-2110-03894} & 93.2 & \cellcolor[HTML]{EFEFEF}{\textbf{95.5 }(\textcolor{myred}{+3.7})} & 90.9 & 79.5 \\ \hhline{~~*{8}{-}} \Tstrut & \Tstrut 
                                               & Arabic SC             &  Ar     & 16                     & 98.9 \cite{DBLP:journals/corr/abs-2110-03894} & 99.7 & \cellcolor[HTML]{EFEFEF}{\textbf{100.0 }(\textcolor{myred}{+1.1}) } & 85.6 & 92.6 \\ 
        \midrule
        IC & ACC ($\uparrow$)                      & Fluent SC             &  En      & 24                     & 99.7 \cite{bermuth2022finstreder} & 97.2 & 97.3 & 98.1 & \cellcolor[HTML]{EFEFEF}{\textbf{98.2 }(\textcolor{mygreen}{-1.5})} \\ 
        \midrule
         LID & ACC ($\uparrow$)                     & Voxforge         &   \makecell{En, Es, Fr \\ De, Ru, It}          & 6                      & 99.8 \cite{shor2022universal} & 90.9 & \cellcolor[HTML]{EFEFEF}
        {\textbf{94.2 }(\textcolor{mygreen}{-5.6})} & 81.8 & 80.4 \\
        \midrule
        FSD  & EER ($\downarrow$)                       & ASVspoof          & En      & 2                      & 2.5 \cite{shor2022universal} & 18.5 & 13.5 & \cellcolor[HTML]{EFEFEF}{\textbf{13.1 }(\textcolor{mygreen}{+10.6})} & 18.3 \\ 
        \midrule
        ER & ACC ($\uparrow$)                       & IEMOCAP          &  En       & 4                      & 79.2 \cite{shor2022universal} & 42.1 & 44.3 & 49.9 & \cellcolor[HTML]{EFEFEF}{\textbf{50.2} (\textcolor{mygreen}{-29})} \\
        \midrule
        AcC & ACC ($\uparrow$)                       & AccentDB          &  En      & 9                      & 99.5 \cite{DBLP:conf/lrec/AhamadAB20} & 78.9 & 83.4 & 86.5 & \cellcolor[HTML]{EFEFEF}{\textbf{87.1} (\textcolor{mygreen}{-12.4})} \\
        \midrule
        \multirow{2}{*}{SD} & \multirow{2}{*}{F1 ($\uparrow$)}        & MUStARD          & En          &  2  & 64.6 \cite{DBLP:conf/acl/CastroHPZMP19} & 55.0   & 77.8     &  74.4        &       \cellcolor[HTML]{EFEFEF}{\textbf{78.7 }(\textcolor{myred}{+13.1})}              \\   \hhline{~~*{8}{-}} \Tstrut & \Tstrut
        & MUStARD++                                               & En      &   2   &   65.2 \cite{DBLP:conf/lrec/RayMNB22}  &     74.0    &  \cellcolor[HTML]{EFEFEF}{\textbf{75.2 }(\textcolor{myred}{+10})}          &  52.7 & 58.2           \\ 
        \midrule
        GID & F1 ($\uparrow$)                       & VoxCeleb1        & En        & 2                      & 98.3 \cite{hechmi2021voxceleb} & 86.2 & 87.3 & \cellcolor[HTML]{EFEFEF}{\textbf{91.6 }(\textcolor{mygreen}{-6.7})} & 86.2 \\
        \midrule
        VAD & ACC ($\uparrow$)                       & \makecell{Google SC v2 \\ \& Freesound}         &  \makecell{En}      & 2                      & 98.8 \cite{jia2021marblenet} & 96.6 & 96.9 & \cellcolor[HTML]{EFEFEF}{\textbf{98.3 }(\textcolor{mygreen}{-0.5})} & 98.1 \\ 
        \midrule
        AuC & ACC ($\uparrow$)                       & ESC-50           &  \xmark       & 50                     & 97.0 \cite{chen2022hts} & 9.0 & \cellcolor[HTML]{EFEFEF}{\textbf{37.5} (\textcolor{mygreen}{-59.5})} & 20.3 & 27.0 \\ 
        \bottomrule  
        \end{tabular}}
    \label{main_result}
\end{table*}

%% file: 4_experiment.tex
\subsection{Tasks and Datasets}
In this paper, we evaluate SpeechPrompt v2 on 10 speech classification tasks from 14 datasets with different languages and speech properties. Following, we briefly describe these tasks
\footnote{The experiment conducted was solely to evaluate model capabilities based on established benchmarks and labels. The established labels don't provide a holistic landscape about gender and emotion. Predicting binary gender based on audio risks misgendering people and is not inclusive of the whole range of gender identities, including nonbinary people. Also, the apparent emotions are not universal, but are instead culturally dependent, and may be expressed in audio differently by different people.}.

\noindent\textbf{Speech Commands Recognition, SCR}: SCR aims to recognize which keyword is presented in a given utterance. We adopted Google Speech Commands~\cite{warden2018speech} and several low-resource datasets in different languages, including Grabo Speech Commands~\cite{renkens2014acquisition}, Lithuanian Speech Commands~\cite{kolesau2020unsupervised}, and Arabic Speech Commands~\cite{benamer2020database}.

\noindent\textbf{Intent Classification, IC}: IC aims to classify utterances into predefined classes to determine the intent of speakers. We used the Fluent Speech Commands dataset~\cite{DBLP:conf/interspeech/LugoschRITB19}, where each utterance has three labels: action, object, and location.

\noindent\textbf{Language Identification, LID}: LID aims to identify the language of an utterance. We used the Voxforge dataset~\cite{maclean2018voxforge} consisting of six different languages for the task.

\noindent\textbf{Fake Speech Detection, FSD}: FSD aims to distinguish real speech from synthetic speech. We used the Logical Access (LA) part of the ASVspoof dataset~\cite{nautsch2021asvspoof}, containing bona fide and spoofed speech.



\noindent\textbf{Emotion Recognition, ER}: ER aims to recognize the emotion for a given utterance. The widely used ER dataset IEMOCAP~\cite{busso2008iemocap} is adopted in this work.

\noindent\textbf{Accent Classification, AcC}: AcC aims to classify different accents of the same language. We used the AccentDB dataset~\cite{DBLP:conf/lrec/AhamadAB20}, containing 4 Indian-English accents, 4 native-English, and 1 metropolitan Indian-English accent.

\noindent\textbf{Sarcasm Detection, SD}: SD aims to detect whether an utterance contains sarcasm. We adopted two datasets, MUStARD~\cite{DBLP:conf/acl/CastroHPZMP19} and MUStARD++~\cite{DBLP:conf/lrec/RayMNB22}. The latter is an extension of the former.

\noindent\textbf{Gender Identification, GID}: GID aims to classify an utterance into genders. We adopted the VoxCeleb1 dataset~\cite{Nagrani17} and obtained the label based on the provided speaker information.

\noindent\textbf{Voice Activity Detection, VAD}: VAD aims to detect whether an input audio contains speech or background noise. We used Google Speech Commands v2~\cite{warden2018speech} as speech data and Freesound dataset~\cite{DBLP:conf/ismir/FonsecaPFFBFOPS17} as background noise data, and mixed the two for evaluating VAD.

\noindent\textbf{Audio Classification, AuC}: AuC aims to classify different environmental sounds across human and nature. We used the ESC-50~\cite{piczak2015esc} with 50 different classes for the task.

\subsection{Implementation Details}
\noindent\textbf{Pre-trained LMs} For prompt tuning on GSLM and pGSLM. We use the pre-trained HuBERT~\cite{DBLP:journals/taslp/HsuBTLSM21} with 100 clusters and its corresponding pre-trained GSLM and pGSLM. The pre-trained speech SSL models, quantizers, and the spoken LMs are available on \emph{fairseq}\footnote{\url{https://github.com/facebookresearch/fairseq/tree/main/examples/textless_nlp}}.

\noindent\textbf{Prompts and Verbalizer} To keep the pipeline simple, we do not search the hyperparameters task by task. For every task, we use prompt length $l = 5$, containing 0.128M parameters. The learnable verbalizer is a simple linear model mapping the discrete units to the classes of each task and contains negligible number of parameters. Overall, the trainable parameters in each task are less than 0.1\% parameters of the spoken LMs ($\sim$150M). For baseline methods, SpeechPrompt with GSLM and pGSLM, we use a random mapping verbalizer instead of the frequency-based verbalizer. We find that the latter does not consistently outperform the former in the preliminary study.

%% file: 5_result.tex
\subsection{Speech Classification Tasks}
Table \ref{main_result} presents the results of SpeechPrompt v2 on a wide range of speech classification tasks. The performances of SpeechPrompt v2 can be grouped into three categories:

\noindent\textbf{(1) Outperform SOTA}: SpeechPrompt v2 achieves new state-of-the-art results on Lithuanian SCR, Arabic SCR, and Sarcasm Detection (SD). The model generalizes well across different languages, demonstrating exceptional performance in low-resource languages such as Lithuanian and Arabic~\cite{DBLP:journals/corr/abs-2110-03894}. With the aid of powerful pre-trained spoken language models, SpeechPrompt v2 also achieves state-of-the-art performance in Sarcasm Detection.

\noindent\textbf{(2) Competitive with SOTA}: SpeechPrompt v2 performs competitively with the SOTA on Google SCR, Grabo SCR, Intent Classification (IC), Language Identification (LID), Gender Classification (GID), and Voice Activity Detection (VAD). Despite experiencing a slight decrease in performance compared to the current SOTA systems in these tasks, SpeechPrompt v2 remains competitive with them. 

\noindent\textbf{(3) Underperform SOTA}: SpeechPrompt v2 underperforms compared to SOTA on Fake Speech Detection (FSD), Emotion Recognition (ER), Accent Classification (AcC), and Audio Classification (AuC). The experiment reveals that SpeechPrompt v2 exhibits poor performance on non-speech and speech with diverse variance.

The experiment also demonstrates the benefits of using learnable verbalizers in SpeechPrompt. With learnable verbalizers, SpeechPrompt with GSLM exhibits consistent performance improvements in the majority of tasks. However, for certain tasks such as Lithuanian SCR and GID, the introduction of learnable verbalizers in pGSLM resulted in performance drops, potentially due to prompt tuning instability. In the next section, we will delve deeper into the reasons why the inclusion of learnable verbalizers is essential in the SpeechPrompt framework.

However, the experiment results also indicate some limitations of SpeechPrompt. Firstly, it suffers from performance drops when dealing with non-speech or speech with diverse variance. This could be because the spoken language models are pre-trained on an English-speech-only corpus. Secondly, the performances with prompt tuning exhibit some variance, as seen in the poor performance of pGSLM on Grabo SC. We hypothesize that the instability of prompt tuning training is the cause of these inconsistencies. To maintain a simple and unified framework, we did not fine-tune any hyperparameters, including prompt length, learning rate, etc., across all experiments presented in Table 1. Therefore, stabilizing prompt tuning remains an area for future work. Possible approaches to address this issue include introducing hypernetworks~\cite{DBLP:journals/corr/abs-2205-12309} and prompt initialization~\cite{DBLP:conf/acl/LiL20}.

Overall, SpeechPrompt demonstrates comparable and even state-of-the-art performance on a wide range of speech classification tasks. Although there are some performance drops in certain tasks, SpeechPrompt offers advantages such as parameter efficiency and a unified feedforward pipeline due to its prompting paradigm. These features make SpeechPrompt v2 a promising alternative to current systems, especially for applications prioritizing efficiency and simplicity.

\begin{figure}[t]
    \centering
    \subfloat[Class ``YES'']{
        \includegraphics[width=\linewidth]{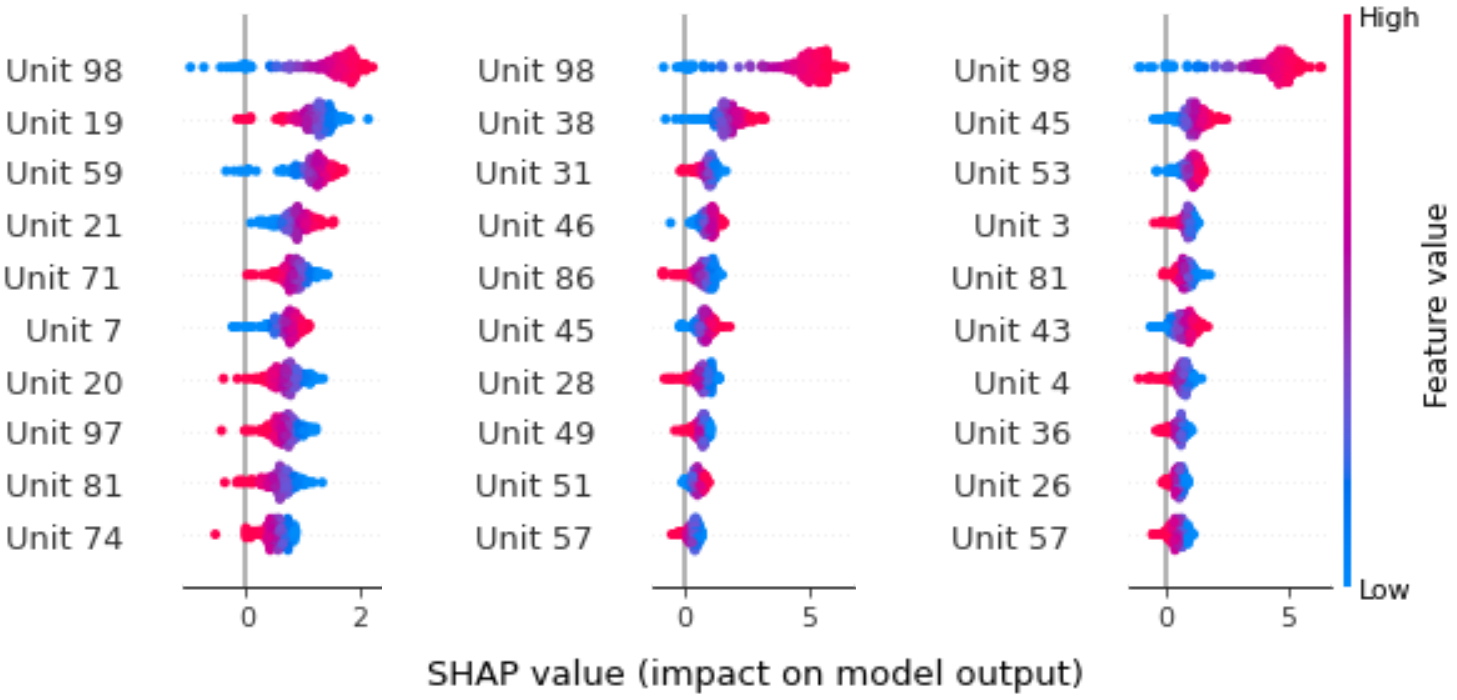}
    }
    
    \subfloat[Class ``LEFT'']{
        \includegraphics[width=\linewidth]{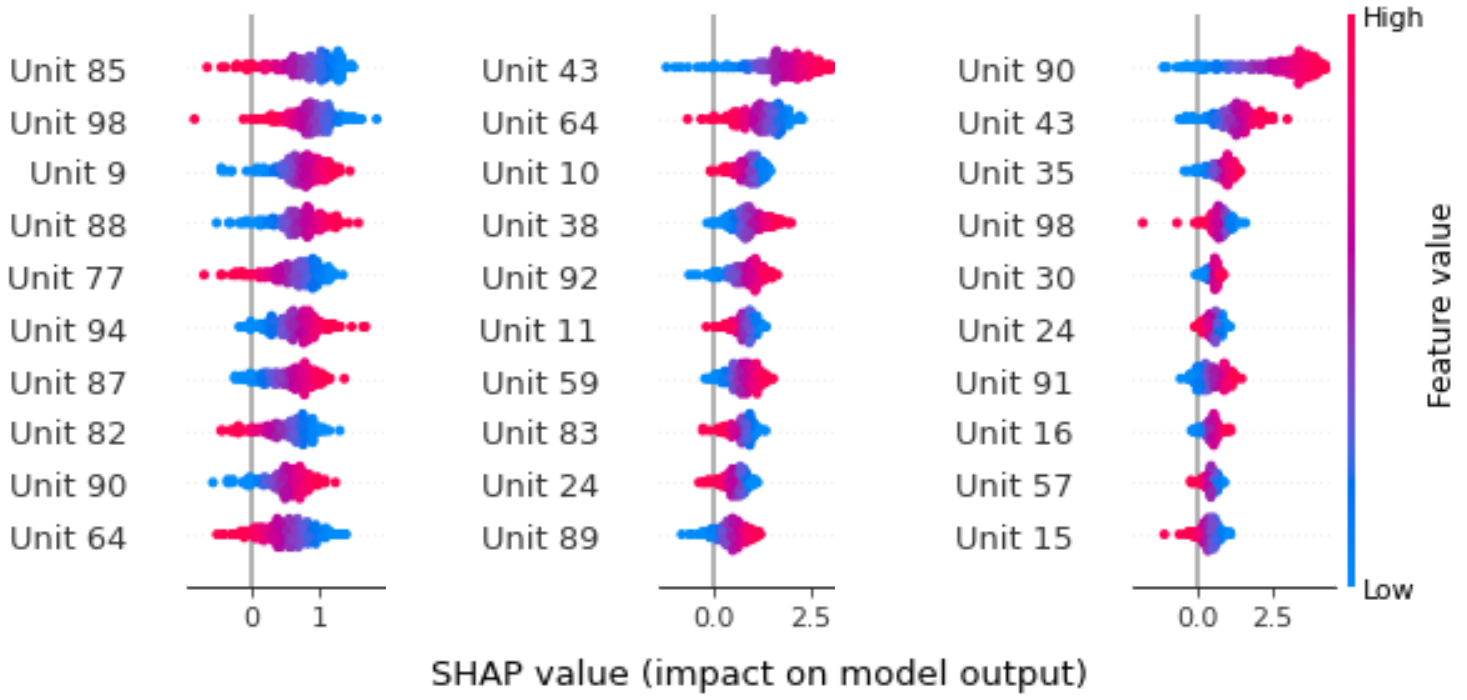}
    }
    \caption{The SHAP analysis of the learnable verbalizer on the Google Speech Command dataset~\cite{vygon2021learning}. Each point refers to one datapoint.} 
    \label{fig:analyzing_verbalizer}
    \vspace{-15pt}
\end{figure}

\vspace{-0.5em}
\subsection{Learnable Verbalizer Analysis}
In this section, we further inspect the behavior of the learnable verbalizer to see how it maps the discrete units to meaningful task labels. Specifically, we adopt Shapley Additive exPlanation~\cite{lundberg2017unified} (SHAP), a popular method to interpret model predictions. In SHAP, the \textbf{SHAP value} is calculated to measure the importance of each input feature, here referring to the discrete unit, to a single prediction made by the model. 

Fig.\ref{fig:analyzing_verbalizer} shows the result of SCR on Google SC dataset for three different runs. We list the top 10 units contributing to the class ``YES'' and the class ``LEFT'' 
\footnote{Please refer to our project website for more demonstration. \url{https://ga642381.github.io/SpeechPrompt}}.
In each chart, the horizontal axis is the SHAP value, while the vertical axis indicates the order of importance of each unit from top to bottom. 
Each point refers to the interpretation of a single datapoint. The color of a point reflects the magnitude of the value for each input unit.
First, we observe that even in the same class, the importance of each unit varies from prediction to prediction. And the learnable verbalizer relies on multiple units to determine the output, indicating a many-to-many mapping behavior from units to labels. This verifies the possible information loss in the previous work, as described in sec\ref{sec:learnable_verbalizer}.
Moreover, the relationship between units and labels does not remain consistent in different runs. 
However, there are units in common and their contribution pattern is similar in different runs (e.g. unit 98 has the same pattern for class ``YES'').
Last, the learnable verbalizer might use one unit to separate different classes. For example, when unit 98 has a lower value, the verbalizer is more likely to predict class ``LEFT'' instead of class ``YES''.
Overall, the analysis suggests that using a learnable verbalizer offers benefits over naive random mapping and frequency-based mapping and is more ideal for our framework.

%% file: 6_conclusion.tex
In this work, we propose SpeechPrompt v2, a prompt tuning framework capable of performing a wide variety of speech classification tasks. We introduce a learnable verbalizer in SpeechPrompt to enhance classification ability. We conducted the experiments with two types of spoken language models, GSLM and pGSLM. The experimental results demonstrated that SpeechPrompt v2 achieves competitive performance and parameter efficiency on different speech classification tasks within a unified framework. We also discussed the limitations of SpeechPrompt, which can guide future research. Overall, we believe our work can inspire the speech community to further explore the potential of the prompting paradigm in speech processing.

%% file: 7_acknowledgement.tex
Part of the work presented here was carried out during the 2022 Jelinek Memorial Summer Workshop on Speech and Language Technologies at Johns Hopkins University, which was supported with unrestricted gifts from Amazon, Microsoft, and Google.